\documentstyle[11pt]{article}
\begin{document}

\title{{\bf The generalized second law for the interacting new agegraphic
dark energy in a non-flat FRW universe enclosed by the apparent
horizon}}
\author{K. Karami\thanks{E-mail: KKarami@uok.ac.ir}\\
A. Abdolmaleki\\\small{Department of Physics, University of
Kurdistan, Pasdaran St., Sanandaj, Iran}\\
}

\maketitle

\begin{abstract}
We investigate the validity of the generalized second law of
gravitational thermodynamics in a non-flat FRW universe containing
the interacting new agegraphic dark energy with cold dark matter.
The boundary of the universe is assumed to be enclosed by the
dynamical apparent horizon. We show that for this model, the
equation of state parameter can cross the phantom divide. We also
present that for the selected model under thermal equilibrium with
the Hawking radiation, the generalized second law is always
satisfied throughout the history of the universe. Whereas, the
evolution of the entropy of the universe and apparent horizon,
separately, depends on the equation of state parameter of the
interacting new agegraphic dark energy model.
\end{abstract}
\clearpage
\section{Introduction}
Type Ia supernovae observational data suggest that the universe is
dominated by two dark components containing dark matter and dark
energy \cite{Riess}. Dark matter (DM), a matter without pressure, is
mainly used to explain galactic curves and large-scale structure
formation, while dark energy (DE), an exotic energy with negative
pressure, is used to explain the present cosmic accelerating
expansion. However, the nature of DE is still unknown, and people
 have proposed some candidates to describe it. The cosmological
 constant, $\Lambda$, is the most obvious theoretical candidate of
 DE, which has the equation of state $\omega=-1$.
 Astronomical observations indicate that the cosmological constant is many orders of magnitude
 smaller than estimated in modern theories of elementary particles \cite{Weinberg}. Also the
 "fine-tuning" and the "cosmic coincidence" problems are the two
 well-known difficulties of the cosmological constant problems
 \cite{Copeland}.

 There are different alternative theories for the dynamical DE scenario which have been
 proposed by people to interpret the accelerating universe. i) The scalar-field models of DE including
 quintessence \cite{Wetterich},
 phantom (ghost) field \cite{Caldwell1}, K-essence \cite{Chiba}
 based on earlier work of K-inflation \cite{Picon3},
 tachyon field \cite{Sen}, dilatonic ghost condensate \cite{Gasperini},
 quintom \cite{Elizalde}, and so forth. ii) The interacting DE models including
 Chaplygin gas \cite{Kamenshchik}, holographic DE models
 \cite{Cohen}, and braneworld models
 \cite{Deffayet}, etc. The interaction between DE and DM has been discussed in ample detail
by \cite{Amendola1}. The recent evidence provided by the Abell
Cluster A586 supports the interaction between DE and DM
\cite{Bertolami}. However, there are no strong observational
bounds on the strength of this interaction \cite{Feng}.

Recently, the original agegraphic dark energy (ADE) and the new
agegraphic dark energy (NADE) models were proposed by Cai
\cite{Cai} and Wei $\&$ Cai \cite {Wei1}, respectively. The
original ADE model cannot explain the matter-dominated era
\cite{Cai}. Thus, Wei and Cai \cite{Wei1} proposed the NADE, while
the age of the universe has been replaced by the conformal time
scale. The evolution behavior of the NADE is very different from
that of original ADE. Instead the evolution behavior of the NADE
is similar to that of the holographic DE \cite{Cohen}. But some
essential differences exist between them. In particular, the NADE
model is free of the drawback concerning causality problem which
exists in the holographic DE model. The ADE and NADE models have
been studied in ample detail by \cite{Wei2,Kim,Sheykhi,Wei3}.

Besides, as usually believed, an early inflation era leads to a
flat universe. This is not a necessary consequence if the number
of e-foldings is not very large \cite{Huang}. It is still possible
that there is a contribution to the Friedmann equation from the
spatial curvature when studying the late universe, though much
smaller than other energy components according to observations.
Therefore, it is not just of academic interest to study a universe
with a spatial curvature marginally allowed by the inflation model
as well as observations. Some experimental data have implied that
our universe is not a perfectly flat universe and that it
possesses a small positive curvature \cite{Bennett}.

In the semiclassical quantum description of black hole physics, it
was found that black holes emit Hawking radiation with a temperature
proportional to their surface gravity at the event horizon and they
have an entropy which is one quarter of the area of the event
horizon in Planck unit \cite{Hawking}. The temperature, entropy and
mass of black holes satisfy the first law of thermodynamics
\cite{Bardeen}. On the other hand, it was shown that the Einstein
equation can be derived from the first law of thermodynamics by
assuming the proportionality of entropy and the horizon area
\cite{Jacobson}. The study on the relation between the Einstein
equation and the first law of thermodynamics has been generalized to
the cosmological context where it was shown that the first law of
thermodynamics on the apparent horizon $\tilde{r}_{\rm A}$ can be
derived from the Friedmann equation and vice versa if we take the
Hawking temperature $T_{\rm A} =1/2\pi \tilde{r}_{\rm A}$ and the
entropy $S_{\rm A} = \pi \tilde{r}_{\rm A}^2$ on the apparent
horizon \cite{Cai05}. Furthermore, the equivalence between the first
law of thermodynamics and Friedmann equation was also found for
gravity with Gauss-Bonnet term and the Lovelock gravity theory
\cite{Cai05,Akbar}.

Besides the first law of thermodynamics, a lot of attention has
been paid to the generalized second law of thermodynamics in the
accelerating universe driven by DE. The generalized second law of
thermodynamics is as important as the first law, governing the
development of the nature
\cite{Gong,Izquierdo2,Izquierdo,Zhau,Sheykhi3,Wang2}. Note that in
\cite{Wang2}, the authors investigated the validity of the first
and the generalized second law of thermodynamics for both apparent
and event horizon for the case of interacting holographic DE with
DM. They showed that in contrast to the case of the apparent
horizon, both the first and second law of thermodynamics breakdown
if one consider the universe to be enveloped by the event horizon
with the usual definitions of entropy and temperature. They argued
that the break down of the first law can be attributed to the
possibility that the first law may only apply to variations
between nearby states of local thermodynamic equilibrium, while
the event horizon reflects the global spacetime properties.
Besides in the dynamic spacetime, the horizon thermodynamics is
not as simple as that of the static spacetime. The event horizon
and apparent horizon are in general different surfaces. The
definition of thermodynamical quantities on the cosmological event
horizon in the nonstatic universe are probably ill-defined. Author
of \cite{Setare3} by redefining the event horizon measured from
the sphere of the horizon as the system's IR cut-off for an
interacting holographic DE model in a non-flat universe, showed
that the second law is satisfied for the special range of the
deceleration parameter. Recently, Sheykhi \cite{Sheykhi2} showed
that both the first and generalized second law of thermodynamics
are respected for the non-flat universe enveloped by the apparent
horizon in braneworld scenarios.

All mentioned in above motivate us to investigate the generalized
second law of thermodynamics for the interacting NADE model with
DM in the non-flat universe enclosed by the apparent horizon with
the Hawking radiation. To do this, in Section 2, we study the NADE
model in a non-flat universe which is in interaction with the cold
DM. In Section 3, we investigate the validity of the generalized
second law of thermodynamics for the present model enclosed by the
apparent horizon which is in thermal equilibrium with the Hawking
radiation. Section 4 is devoted to conclusions.
\section{Interacting NADE model and DM}

We consider the Friedmann-Robertson-Walker (FRW) metric for the
non-flat universe as
\begin{equation}
{\rm d}s^2=-{\rm d}t^2+a^2(t)\left(\frac{{\rm
d}r^2}{1-kr^2}+r^2{\rm d}\Omega^2\right),\label{metric}
\end{equation}
where $k=0,1,-1$ represent a flat, closed and open FRW universe,
respectively. Observational evidences support the existence of a
closed universe with a small positive curvature ($\Omega_{k}\sim
0.02$) \cite{Bennett}. Define $\tilde{r} = ar$, the metric
(\ref{metric}) can be rewritten as ${\rm d}s^2 = h_{ab}{\rm
d}x^a{\rm d}x^b + \tilde{r}^2{\rm d}\Omega^2$, where $x^a = (t, r)$,
$h_{ab}$ = diag($-1, a^2/(1 - kr^2)$). By definition,
$h^{ab}\partial_{a}\tilde{r}\partial_{b}\tilde{r} = 0$, the location
of the apparent horizon in the FRW universe is obtained as
$\tilde{r} = r_{\rm A} = (H^2+k/a^2)^{-1/2}$ \cite{Cai09}. For $k =
0$, the apparent horizon is same as the Hubble horizon.

The first Friedmann equation for the non-flat FRW universe has the
following form
\begin{equation}
{\textsl{H}}^2+\frac{k}{a^2}=\frac{8\pi}{3}~
(\rho_{\Lambda}+\rho_{\rm m}),\label{eqfr}
\end{equation}
where we take $G=1$. Also $\rho_{\Lambda}$ and $\rho_{\rm m}$ are
the energy density of DE and DM, respectively. Let us define the
dimensionless energy densities as
\begin{equation}
\Omega_{\rm m}=\frac{\rho_{\rm m}}{\rho_{\rm
cr}}=\frac{8\pi\rho_{\rm m}}{3H^2},~~~~~~\Omega_{\rm
\Lambda}=\frac{\rho_{\Lambda}}{\rho_{\rm
cr}}=\frac{8\pi\rho_{\Lambda}}{3H^2},~~~~~~\Omega_{k}=\frac{k}{a^2H^2},
\label{eqomega}
\end{equation}
then, the first Friedmann equation yields
\begin{equation}
\Omega_{\rm m}+\Omega_{\Lambda}=1+\Omega_{k}.\label{eq10}
\end{equation}
Following \cite{Sheykhi}, the energy density of the NADE is given by
\begin{equation}
\rho_{\Lambda}=\frac{3{n}^2}{8\pi\eta^2},\label{NADE}
\end{equation}
where $n$ is a constant. The astronomical data gives $n =
2.716_{-0.109}^{+0.111}$ \cite{Wei3}. Also $\eta$ is conformal
time of the FRW universe, and given by
\begin{equation}
\eta=\int\frac{{\rm d}t}{a}=\int_{0}^{a}\frac{{\rm d}a}{Ha^2}.
\end{equation}
The DE density (\ref{NADE}) is similar to that of the holographic
DE \cite{Cohen}, but the conformal time stands instead of the
future event horizon distance of the universe. This solves the
causality problem which exists in the holographic DE model.
Because the existence of the future event horizon requires an
eternal accelerated expansion of the universe \cite{Wei1}.

From definition $\rho_{\Lambda}=3H^2\Omega_{\Lambda}/8\pi$, we get
\begin{equation}
\eta=\frac{{n}}{H\sqrt{\Omega_{\Lambda}}}.\label{eta}
\end{equation}

We consider a universe containing an interacting NADE density
$\rho_{\Lambda}$ and the cold dark matter (CDM), with $\omega_{\rm
m}=0$. The energy equations for NADE and CDM are given by
\begin{equation}
\dot{\rho}_{\Lambda}+3H(1+\omega_{\Lambda})\rho_{\Lambda}=-Q,\label{eqpol}
\end{equation}
\begin{equation}
\dot{\rho}_{\rm m}+3H\rho_{\rm m}=Q,\label{eqCDM}
\end{equation}
where following \cite{Kim06}, we choose $Q=\Gamma\rho_{\Lambda}$ as
an interaction term and
$\Gamma=3b^2H(\frac{1+\Omega_{k}}{\Omega_{\Lambda}})$ is the decay
rate of the NADE component into CDM with a coupling constant $b^2$.
 The choice of the interaction
between both components was to get a scaling solution to the
coincidence problem such that the universe approaches a stationary
stage in which the ratio of DE and DM becomes a constant
\cite{Hu}. The dynamics of interacting DE models with different
$Q$-classes have been studied in ample detail by \cite{Pavon}. The
freedom of choosing the specific form of the interaction term $Q$
stems from our incognizance of the origin and nature of DE as well
as DM. Moreover, a microphysical model describing the interaction
between the dark components of the universe is not available
nowadays \cite{Setare2}.

Taking the time derivative of Eq. (\ref{NADE}), using
$\dot{\eta}=1/a$ and Eq. (\ref{eta}) yields
\begin{equation}
\dot{\rho}_{\Lambda}=-\frac{2H\sqrt{\Omega_{\Lambda}}}{{
n}a}\rho_{\Lambda}.\label{rhodot}
\end{equation}
Substituting Eq. (\ref{rhodot}) in (\ref{eqpol}), gives the
equation of state (EoS) parameter as
\begin{eqnarray}
\omega_{\Lambda}=-1+\frac{2\sqrt{\Omega_{\Lambda}}}{3{
n}a}-b^2\Big(\frac{1+\Omega_{k}}{\Omega_{\Lambda}}\Big),\label{w}
\end{eqnarray}
which shows that for $b^2 = 0$ then $\omega_{\Lambda}>-1$ and
cannot behaves phantom EoS. However, in the presence of
interaction, $b^2\neq 0$, taking $\Omega_{\Lambda}=0.72$,
$\Omega_k=0.02$ \cite{Bennett}, $n=2.7$ \cite{Wei3} and $a=1$ for
the present time, Eq. (\ref{w}) gives
\begin{equation}
\omega_{\Lambda}=-0.79-1.42b^2,\label{w1}
\end{equation}
which clears that the phantom EoS $\omega_{\Lambda}<-1$ can be
obtained when $b^2>0.15$ for the interaction between NADE and CDM.

The deceleration parameter is given by
\begin{equation}
q=-\Big(1+\frac{\dot{H}}{H^2}\Big).\label{q1}
\end{equation}
Taking the time derivative in both sides of Eq. (\ref{eqfr}), and
using Eqs. (\ref{eqomega}), (\ref{eq10}), (\ref{eqpol}) and
(\ref{eqCDM}), we get
\begin{equation}
q=-\frac{3}{2}\Omega_{\Lambda}+\frac{\Omega_{\Lambda}^{3/2}}{na}+\frac{1}{2}(1-3b^2)(1+\Omega_{k}).\label{q3}
\end{equation}
Now taking $\Omega_{\Lambda}=0.72$, $\Omega_k=0.02$
\cite{Bennett}, $n=2.7$ \cite{Wei3} and $a=1$ for the present time
we get
\begin{equation}
q=-0.34-1.53b^2,\label{q4}
\end{equation}
which is always negative even in the absence of interaction
between NADE and CDM. Therefore the NADE model in the present time
can drive the universe to accelerated expansion.

\section{Generalized second law of thermodynamics}
Here, we study the validity of the generalized second law (GSL) of
gravitational thermodynamics. According to the GSL, entropy of the
NADE and CDM inside the horizon plus the entropy of the horizon do
not decrease with time \cite{Wang2}. In the GSL, the definition of
the temperature of the fluid is important. Cai \& Kim \cite{Cai05}
proofed that the Friedmann equations in Einstein gravity are derived
by applying the first law of thermodynamics to the dynamic apparent
horizon, $\tilde{r}_{\rm A}=(H^2+k/a^2)^{-1/2}$, of a FRW universe
with any spatial curvature in arbitrary dimensions and assuming that
the apparent horizon has an associated entropy $S_{\rm A}$ and
Hawking temperature $T_{\rm A}$ as $S_{\rm A} = \pi \tilde{r}_{\rm
A}^2$, $T_{\rm A} =1/2\pi \tilde{r}_{\rm A}$. In the braneworld
scenarios, the Friedmann equations also can be written directly in
the form of the first law of thermodynamics, at the apparent horizon
with the Hawking temperature on the brane, regardless of whether
there is the intrinsic curvature term on the brane or a Gauss-Bonnet
term in the bulk \cite{Sheykhi2}. Recently the Hawking radiation
with temperature $T_{\rm A} = 1/2\pi \tilde{r}_{\rm A}$ on the
apparent horizon of a FRW universe with any spatial curvature has
been observed in \cite{Cai09} which gives more solid physical
implication of the temperature associated with the apparent horizon.
The Hawking temperature is measured by an observer with the Kodoma
vector inside the apparent horizon \cite{Cai09}. These motivate us
to consider the Hawking temperature of the dynamic apparent horizon
for our model. We also limit ourselves to the assumption that the
thermal system including the NADE and CDM bounded by the apparent
horizon remain in equilibrium so that the temperature of the system
must be uniform and the same as the temperature of its boundary.
This requires that the temperature $T$ of the both NADE and CDM
inside the apparent horizon should be in equilibrium with the
Hawking temperature $T_{\rm A}$ associated with the apparent
horizon, so we have $T = T_{\rm A}$. This expression holds in the
local equilibrium hypothesis. If the temperature of the system
 differs much from that of the horizon, there will be spontaneous heat flow between
  the horizon and the fluid and the local equilibrium hypothesis will no
longer hold \cite{Izquierdo,Zhau}.

The entropy of the universe including the DE and CDM inside the
apparent horizon can be related to its energy and pressure in the
horizon by Gibb's equation \cite{Izquierdo,Zhau,Wang2}
\begin{equation}
T{\rm d}S={\rm d}E+P{\rm d}V,\label{eqSLT1}
\end{equation}
where like \cite{Wang2}, ${\rm V}=4\pi \tilde{r}_{\rm A}^3/3$ is
the volume containing the NADE and CDM with the radius of the
apparent horizon $\tilde{r}_{\rm A}$ and $T=T_{\rm A}=1/(2\pi
\tilde{r}_{\rm A})$ is the Hawking temperature of the apparent
horizon. Also
\begin{equation}
E=\frac{4\pi \tilde{r}_{\rm A}^3}{3} (\rho_{\Lambda}+\rho_{\rm
m}),\label{eqEde}
\end{equation}
\begin{equation}
P=P_{\Lambda}+P_{\rm
m}=P_{\Lambda}=\omega_{\Lambda}\rho_{\Lambda}=\frac{3H^2}{8\pi}\omega_{\Lambda}\Omega_{\Lambda}.
\label{eqEcdm}
\end{equation}
For the dynamical apparent horizon
\begin{equation}
\tilde{r}_{\rm A}=H^{-1}(1+\Omega_{k})^{-1/2},\label{ah}
 \end{equation}
its derivative with respect to cosmic time $t$ yields
\begin{equation}
\dot{\tilde{r}}_{\rm
A}=\frac{3(1+\Omega_{k}+\Omega_{\Lambda}\omega_{\Lambda})}{2(1+\Omega_{k})^{3/2}}\label{ahdot}.
\end{equation}
Taking the derivative in both sides of (\ref{eqSLT1}) with respect
to cosmic time $t$, and using Eqs. (\ref{eqfr}), (\ref{eqomega}),
(\ref{eq10}), (\ref{eqpol}), (\ref{eqCDM}), (\ref{eqEde}),
(\ref{eqEcdm}), (\ref{ah}), and (\ref{ahdot}), we obtain the
evolution of the entropy in the universe containing the NADE and
CDM as
\begin{eqnarray}
\dot{S}=\frac{3\pi}{2H(1+\Omega_{k})^3}(1+\Omega_{k}+3\Omega_{\Lambda}\omega_{\Lambda})
(1+\Omega_{k}+\Omega_{\Lambda}\omega_{\Lambda}).\label{Sah}
\end{eqnarray}
Equation (\ref{Sah}) shows that for
\begin{equation}
\omega_{\Lambda}\leq-\Big(\frac{1+\Omega_{k}}{\Omega_{\Lambda}}\Big),~~~~~~
\omega_{\Lambda}\geq-\frac{1}{3}\Big(\frac{1+\Omega_{k}}{\Omega_{\Lambda}}\Big),\label{wq}
\end{equation}
the contribution of the entropy of the universe inside the dynamical
apparent horizon in the GSL is positive, i.e. $\dot{S}\geq0$. Taking
$\Omega_{\Lambda}=0.72$, $\Omega_k=0.02$ \cite{Bennett} for the
present time and using Eq. (\ref{w1}), conditions (\ref{wq}) reduce
to $b^2\geq0.44$.

Also in addition to the entropy in the universe, there is a
geometric entropy on the apparent horizon $S_{{\rm A}}=\pi
\tilde{r}_{\rm A}^{2}$ \cite{Wang2}. The evolution of this horizon
entropy is obtained as
\begin{eqnarray}
\dot{S}_{\rm
A}=\frac{3\pi}{H(1+\Omega_{k})^2}(1+\Omega_{k}+\Omega_{\Lambda}\omega_{\Lambda}).\label{SAah}
\end{eqnarray}
Equation (\ref{SAah}) clears that for
\begin{equation}
\omega_{\Lambda}\geq-(\frac{1+\Omega_{k}}{\Omega_{\Lambda}}),\label{wf}
\end{equation}
the contribution of the dynamical apparent horizon in the GSL is
positive, i.e. $\dot{S}_{\rm A}\geq 0$. Taking the above-mentioned
values for the fractional densities for the present time and using
Eq. (\ref{w1}) again, condition (\ref{wf}) yields $b^2\leq 0.44$.
Therefore for a given coupling constant of interaction $b^2$ at the
present time, entropy of the universe and apparent horizon can not
be an increasing function of time, simultaneously.

Finally, using Eqs. (\ref{Sah}) and (\ref{SAah}), the GSL due to
different contributions of the NADE, CDM and apparent horizon can
be obtained as
\begin{equation}
\dot{S}_{\rm
tot}=\frac{9\pi}{2H(1+\Omega_{k})^3}(1+\Omega_{k}+\Omega_{\Lambda}\omega_{\Lambda})^2\geq
0,\label{Stotah}
\end{equation}
where $S_{\rm tot}=S+S_{\rm A}$ is the total entropy. Equation
(\ref{Stotah}) presents that the GSL for the universe containing
the interacting NADE with CDM enclosed by the dynamical apparent
horizon is always satisfied throughout the history of the universe
for any spatial curvature and it is independent of the EoS
parameter of the interacting NADE model.
\section{Conclusions}
Here we considered the NADE model, in the presence of interaction
between DE and DM, for the universe with spatial curvature. We
obtained the EoS for interacting NADE density in a non-flat
universe. In the presence of interaction between NADE and CDM, the
EoS parameter of NADE, $\omega_{\Lambda}$, behaves like phantom DE
in the non-flat FRW universe.

We assumed that the universe to be in thermal equilibrium with the
Hawking temperature on the apparent horizon. The apparent horizon is
important for the study of cosmology, since on the apparent horizon
there is the well known correspondence between the first law of
thermodynamics and Einstein equation. We found that for a non-flat
universe enclosed by the apparent horizon with the Hawking radiation
and containing an interacting NADE with CDM, the generalized second
law of gravitational thermodynamics is always respected for any
spatial curvature, independently of the EoS parameter of the
interacting NADE model. Whereas, the evolution of the entropy of the
universe and apparent horizon, separately, depends on the EoS
parameter of the model.

\end{document}